# Polyvinyl Alcohol-Few Layer Graphene Composite Films Prepared from Aqueous Colloids. Investigations of Mechanical, Conductive and Gas Barrier Properties

**Benoit Van der Schueren [1], Hamza El Marouazi [1], Anurag Mohanty [1], Patrick Lévêque [2], Christophe Sutter [1], Thierry Romero [1] and Izabela Janowska [1,*]**

1 Institut de Chimie et Procédés pour l'Énergie, l'Environnement et la Santé (ICPEES), CNRS UMR 7515-University of Strasbourg, 25 rue Becquerel, 67087 Strasbourg, France; benelux88@gmail.com (B.V.d.S.); hamza.el-marouazi@etu.unistra.fr (H.E.M.); anurag.mohanty@etu.unistra.fr (A.M.); christophe.sutter@unistra.fr (C.S.); thierry.romero@unistra.fr (T.R.)

2 Laboratoire des sciences de l'Ingénieur, de l'Informatique et de l'Imagerie (ICube), UMR 7357, CNRS, Université de Strasbourg, 67400 Strasbourg, France; patrick.leveque@unistra.fr

* Correspondence: janowskai@unistra.fr

**Abstract:** Quasi all water soluble composites use graphene oxide (GO) or reduced graphene oxide (rGO) as graphene based additives despite the long and harsh conditions required for their preparation. Herein, polyvinyl alcohol (PVA) films containing few layer graphene (FLG) are prepared by the co-mixing of aqueous colloids and casting, where the FLG colloid is first obtained via an efficient, rapid, simple, and bio-compatible exfoliation method providing access to relatively large FLG flakes. The enhanced mechanical, electrical conductivity, and $O_2$ barrier properties of the films are investigated and discussed together with the structure of the films. In four different series of the composites, the best Young's modulus is measured for the films containing around 1% of FLG. The most significant enhancement is obtained for the series with the largest FLG sheets contrary to the elongation at break which is well improved for the series with the lowest FLG sheets. Relatively high one-side electrical conductivity and low percolation threshold are achieved when compared to GO/rGO composites (almost $10^{-3}$ S/cm for 3% of FLG and transport at 0.5% FLG), while the conductivity is affected by the formation of a macroscopic branched FLG network. The composites demonstrate a reduction of $O_2$ transmission rate up to 60%.

**Keywords:** nano composites; polymer-matrix composites; electrical properties; mechanical properties; few layer graphene composites

## 1. Introduction

Most of the polymer-graphene composites contain graphene oxide (GO) or graphene platelets, usually prepared by exfoliation in organic solvent. In case of water-soluble polymers such as PVA, GO would be the first choice due to the high content of hydrophilic/oxygenated groups and related ability to be well dispersed in water. The second advantage is usually a high aspect ratio, i.e., high lateral size of sheets vs. low thickness. However, the required reduction to restore conjugated C=C lattice and harsh synthesis conditions are important drawbacks of GO [1,2]. Finally, the reduced GO (rGO) is much less dispersible in water and some reports include additional functionalization of rGO to mix and get higher interface with PVA [3–6]. High strength and ductility composites were obtained for instance using rhodamine to functionalize rGO by cation-π and π-π interactions with an rGO surface [5]. The interesting mechanical and conductive properties were achieved for the composites with self-assemblies of microscale PVA lamellas, dendrites, and rods containing rGO functionalized by sulfonate groups [6].

The common way to obtain GO/rGO involves harsh conditions via the Hummer method, and reduction with hydrazine followed by preparation of the composite [7–11]. Few other reports deal for instance with the use of stress induced exfoliation to graphene based on direct ultrasonication of graphite with PVA. However, this latter method is long as well (48 h of sonication), and the properties of the final composites are not spectacular, as eventually, additional drawings to aligned graphene are applied [12].

Since GO is much more dispersible in water, graphene/rGO based composites show in general superior mechanical and conductive properties. The comparative studies between GO and rGO have been done for instance by Bao et al. [7]. The work clearly shows the advantage of graphene (rGO) over GO in PVA composites with related mechanisms in term of mechanical, conductivity and thermal stability properties [7]. On the other hand, a study demonstrated that, due to the presence of the basal oxygen containing groups and consequently adequate interfacial interactions with PVA, GO containing composite (0.3%) showed higher tensile strength than the one containing rGO [9].

Most reported works deal with low graphene loading and an interesting enhancement of mechanical, electrical or barrier properties were obtained at low graphene (rGO) content: 0.5–0.8% [7,11,13,14]. Some improvements were observed at much lower concentrations [4,15]. The best results in term of tensile strength were obtained by Zhao for the composite containing rGO, obtained in the presence of surfactant Sodium Dodecyl Benzene Sulphonate (SDBS), showing a 1000% enhancement of Young's modulus for 1.8% rGO content. However, no conductivity properties were demonstrated [8].

Despite a number of studies reporting on GO and rGO additives, the use of FLG/multi-layer graphene (MLG) for the same purpose is rare [16,17]. This is probably due to the lack of FLG that would fulfil the requirement of efficient and simple synthesis with a relatively high aspect ratio of the product. Most recently synthetized bio-compatible FLGs, or graphene platelets differently speaking, do not exceed a few hundred nm in size [18]. According to the experimental and theoretical evaluations, FLG/MLG has a similar Young's modulus, i.e., around 1TPa, to monolayer graphene [19], while the conductivity is affected by the perpendicular component that is three orders of magnitude lower than in the planar direction [20].

The aim of this work is to investigate FLG-containing PVA films obtained by casting from the co-mixed aqueous colloids, where FLG colloid are previously prepared by rapid, efficient and simple mixing assisted-exfoliation of expanded graphite in the presence of bovine serum albumin: a natural system possessing a high hydrophilic/lipophilic balance (HLB) [21,22]. The resulting FLG has a relatively high aspect ratio due to the large size of the sheets that is comparable with some reported GO/rGO sheets with lower size, while the simplicity of its preparation is a significant improvement on the harsh conditions typically applied for GO/rGO synthesis. On the other hand, the number of sheets is clearly more important. It is then of interest to check the possibility to use such FLGs and their water processable colloid, and compare the properties of the obtained composites with those known in the literature.

## 2. Material and Methods

### *2.1. Materials*

Few layer graphene (FLG) was synthesized according to the exfoliation method described earlier [21]: expanded graphite (EG) was submitted to the mixing-assisted ultrasonication in distilled water in the presence bovine serum albumin (BSA) with 5–10 wt. % vs. FLG. In this approach, three different durations of the sonication were applied, i.e., 2, 4, or 6 h. The aqueous colloids obtained after sonication were left for decantation of 4 h and then the stable colloidal supernatants were separated and used later for the preparation of the composites.

EG was purchased from Mersen France Gennevilliers (Gennevilliers, France).

BSA was purchased from Merck (Sigma-Aldrich) company, St. Quentin Fallavier, France.

PVA powder (M.W. 57,000–66,000) was purchased from Alfa Aesar (Kandel, Germany).

For the preparation of composite films (PVA-FLG I, PVA-II, PVA-III, PVA-IV), PVA powder was first dissolved in distilled water by mixing and heating of the solution at 90 °C for 3–6 h (6.25 g/150 mL). Then, the given volume of aqueous suspension containing appropriate weight of FLG were added to the solution. All was next mixed for additional 0.5h and casted in Petri dishes. The casted colloids were dried under ambient conditions for 24 h and next under vacuum at 40–60 °C (or directly in oven at 60 °C in the case of PVA-FLG I).

Additional drying for 48 h at 60 °C was applied prior to the mechanical, conductive, and barrier properties measurements.

### 2.2. Characterization Tools

Thermogravimetric analysis (TGA) was performed on TA instrument SDT Q600 (Luken's drive, New Castle, DE, USA) under $N_2$ and air flow with the heating rate of 15°/min.

Differential scanning calorimetry (DSC) was carried out on Q200 V24.11 Build 124 (USA) for two cycles under $N_2$ with ramp rate of heating and cooling of 10 and 5 °C/min respectively, and isothermal steps of 2 min.

Scanning electron microscopy (SEM) was performed on Jeol JSM-6700 F (Tokyo, Japan) working at 3 kV accelerated voltage. In order to analyze the cross-section some films were broken (or cut) after freezing in liquid $N_2$.

X-ray photoelectron spectroscopy (XPS) was performed on a MULTILAB 2000 Thermo Fisher Scientific (Dreieich, Germany) spectrometer equipped with Al K anode (hν $\frac{1}{4}$ 1486.6 eV). CASA XPS program with a Gaussian-Lorentzian mix function and Shirley background subtraction were employed for processing of spectra.

Transmission electron microscopy (TEM) was performed on 2100 F Jeol microscope (Tokyo, Japan).

Tensile testing was carried out on INSTRON ElectroPuls E3000 (Norwood, MA, USA). Prior to the measurements the composites films were cut into specimens with appropriate dimensions for using in serrated grips.

Conductivity measurements (sheet resistance) of the films were carried out using four points probes method (FPPs) under $N_2$ atmosphere (in glove-box) on KEITHLEY 4200-SCS (Beaverton, OR, USA).

Oxygen transmission analysis was performed on apparatus module Ox-Tran 2/10 (Mocon, USA). For each film two measurements were carried out in accordance with standard ASTM D 3985–05. The samples were first conditioned and next measured for at least 24 h at t = 23 °C and RH = 0%.

## 3. Results and Discussion

### 3.1. Synthesis and Structures of FLG and PVA-FLG

Four series of PVA-FLG composites containing up to 3% of FLG are prepared and denoted as PVA-FLG I, PVA-FLG II, PVA-FLG III, and PVA-FLG IV. The series I and III include FLG that was exfoliated for 2 h (FLG-2h), series II includes FLG obtained after 4 h of exfoliation (FLG-4h), while series IV contains FLG after 6 h of exfoliation (FLG-6h). According to our earlier observations, the duration of exfoliation/ultrasonication of expanded graphite influences the size of FLG flakes which successively decreases with the increase of sonication duration [21]. Additionally, the final drying of PVA-FLG I film was different from the one applied in the three others. Different average size of FLG sheets in series could be observed through SEM microscopy, as shown in Figure 1a,b. Figure 1c shows an additional representative micrograph of a global view of FLG-2h flakes with average size of 5 μm. The prolonged time of sonication also causes the introduction of defects that are mostly linear defects, i.e., edges (due to the lower size of flakes) which is reflected by a lower combustion temperature, as shown in Figure 1d, and enlarged C1s XPS spectra (full width at half-maximum, FWHM), shown in Figure 1e. It is known that, contrary to well crystallized planar conjugated $sp^2$ C=C lattice, graphene/FLG edges contain unsaturated carbon atoms and consequently are more sensitive towards oxidation and greedy for more electronegative oxygen, nitrogen-rich groups. The relative content of O and N, Figure 1e, increases with sonication time, where its eventual negligible variation possibly has origins in the variable amount of stabilizing BSA. A different stabilization behavior in the FLG-BSA suspensions during the decantation step and so different amount of BSA remaining in the suspensions would be a consequence of vary FLG flake size. Figure 1f illustrates additionally deconvoluted C1s XPS spectra of FLG-6h, confirming the presence of very few commonly recognized oxygen-containing groups and $sp^3$ type defect-linked C [23]. According to our current and earlier observations, these types of groups are mostly localized at the edges of FLG and not at the surface as is the case of GO or weakly reduced GO with a number of hydroxyl and epoxy surface-localized groups [21].

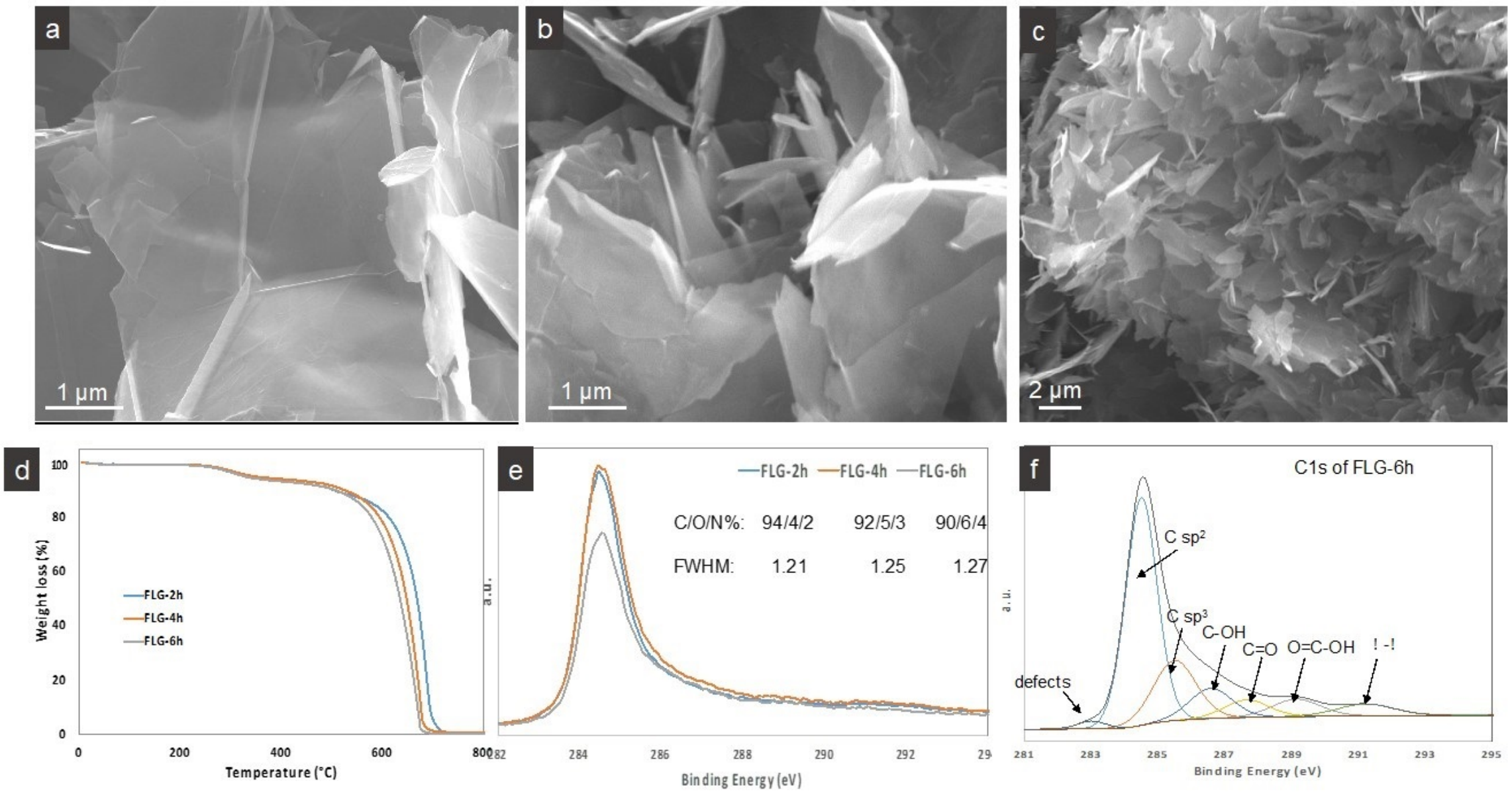

**Figure 1.** Representative SEM micrographs of (**a**,**c**) FLG-2h (**b**) FLG-6 h. (**d**,**e**) TGA curves and XPS spectra of FLG-2h, FLG-4h, FLG-6h, (**f**) deconvoluted XPS spectra of FLG-6h.

The prepared films are the composites of PVA, FLG and very small quantities of BSA present in the amount of 5–10% of FLG. The content of BSA is then very low but sufficient to get a homogenous aqueous dispersion of FLG and next of PVA-FLG (for 1% of FLG vs. weight of PVA, it represents only 0.05–0.1%). BSA is a large system consisting of transport type proteins containing almost 600 amino acid residues, i.e., a huge number of groups able to support hydrogen bonding [24]. Some of these groups are basically occupied to form folded quaternary structures and other ones are responsible for the very good solubility in water. Once submitted to the strong ultrasonication in water, the sonolysis process occurs and BSA form large microspheres through the interactions of disulphide from cysteine residue [21]. According to our observations, during the sonolysis of BSA in the presence of FLG smaller BSA spheres were formed from the excess of BSA [21], while unfolded linear structures of BSA were adsorbed on FLG surface [25]. Concerning the interactions with graphene surfaces, it was somehow reported that the highest affinity in BSA towards graphite surfaces have peptide groups [26]. This time, we could have localized adsorbed BSA and show that the chains of BSA are adsorbed at the surface of the FLG flakes as well as locally at the FLG edges (Supplementary Materials, Figure S1). The adsorption on the edges is not surprising taking into account the existence of reactive groups from the edges that can easily form hydrogen bonding with BSA. It seems also that BSA partially exists in aggregate form and partially as a thin, more planar structure adsorbed at FLG surface, but detailed investigations are necessary for the determination of this structure, which is out of scope of this work, especially given that BSA did not affect the properties studied below (PVA + BSA).

Hydrogen bonds are indeed the central interactions to discuss. They allow first to overcome van der Waals interactions between the sheets in graphite to produce FLG flakes and keep the hydrophobic FLG flakes as a stable aqueous suspension, and further, to make aqueous suspension with PVA. Concerning PVA-FLG interactions, most hydrogen bonding is established due to the presence of BSA adsorbed over FLG and not by oxygen groups covalently bonded to the surface as the case of GO or rGO sheets. Still, other interactions can occur with naked, BSA-free edges of FLG flakes. The role of water in the formation of intermediate hydrogen bonds cannot be forgotten, since possibly affect the properties of the composites. The importance of water in hydrogen bonding formation was demonstrated clearly by Brinson et al., who proposed a gallery-bridging hydrogen bond network associated with rotational and translational degrees of freedom thanks to water molecules in filtered GO paper and its PVA composite. The gallery played a role in the transfer of stress and modification of the stiffness [27].

The common feature of all films is the formation a macroscopic branched network of FLG with dendritic features at the periphery, visible from the upper side. The network could be observed during the drying step of casted colloids, Figure 2a. We associate this phenomenon with the diffusion limited aggregation (DLA) process induced by drying, quite often present in colloid science. A similar kind of branched pattern was also previously obtained upon the drying of pure FLG dispersed in organic solvents [28]. The DLA process and resulting patterns depend on many parameters related to the chemical potential of the matter, its concentration, overall interactions between the matter, matter-solvent and matter-support, matter-air, and drying conditions. It seems then that the present system including PVA, FLG covered by BSA, and water under casting/drying conditions is favorable for DLA to occur (for the reason of comparison FLG-BSA-PLA film prepared in parallel under the same drying conditions did not reveal the formation of such network). Some tendency to form a kind of wavy network is observed in ref. PVA film (Supplementary Materials Figure S2), while in the composites, the branches of the network are full of arranged FLG flakes as we could have confirmed by low resolution SEM analysis (Supplementary Materials Figure S3). Figure 2b–e shows the representative SEM micrographs of PVA-FLG films. Two sides of films can be distinguished after their drying, the one having an opaque optical aspect, that being upper side, seen in Figure 2b, and the one shining more, being in contact with glass (or plastic) Petri dish under drying, shown in Figure 2c. In Figure 2d,e, the transversal cross-section of the films shows the FLG flakes embedded in, and locally protruded, from polymer matrix. We can see clearly the parallel to the film surface (planar) orientation of the FLG flakes within the composite, the tendency observed in the literature rather for graphene (rGO) than for GO [9]. It was indeed observed that, due to the presence of the basal oxygen containing groups, and consequently adequate interfacial interactions with PVA, GO was randomly arranged within PVA matrix contrary to rGO that demonstrate more aligned arrangement [9]. According to this statement, the planar arrangement of FLG flakes suggest that the interactions between FLG edges (free or BSA decorated) and PVA occur and are relatively significant. Some deviation from planar arrangement was observed only close to the upper side of the film.

The TEM micrographs in Figure 3 confirm the existence of few micrometer size FLG flakes having good interface with the polymer, as well as few sheets in the flakes, i.e., nine in the presented micrograph of Figure 3c. According to our earlier studies the estimation of the average number of the sheets in FLG used here is around seven, while the presence of up to 15 as well as 2, 3 sheets are observed by TEM microscopy. Such number of sheets in FLG flakes seems to be high compared to GO/rGO materials but the thickness of the flakes is counterbalanced by relatively large size of the flakes, i.e., 3–5 µm on average.

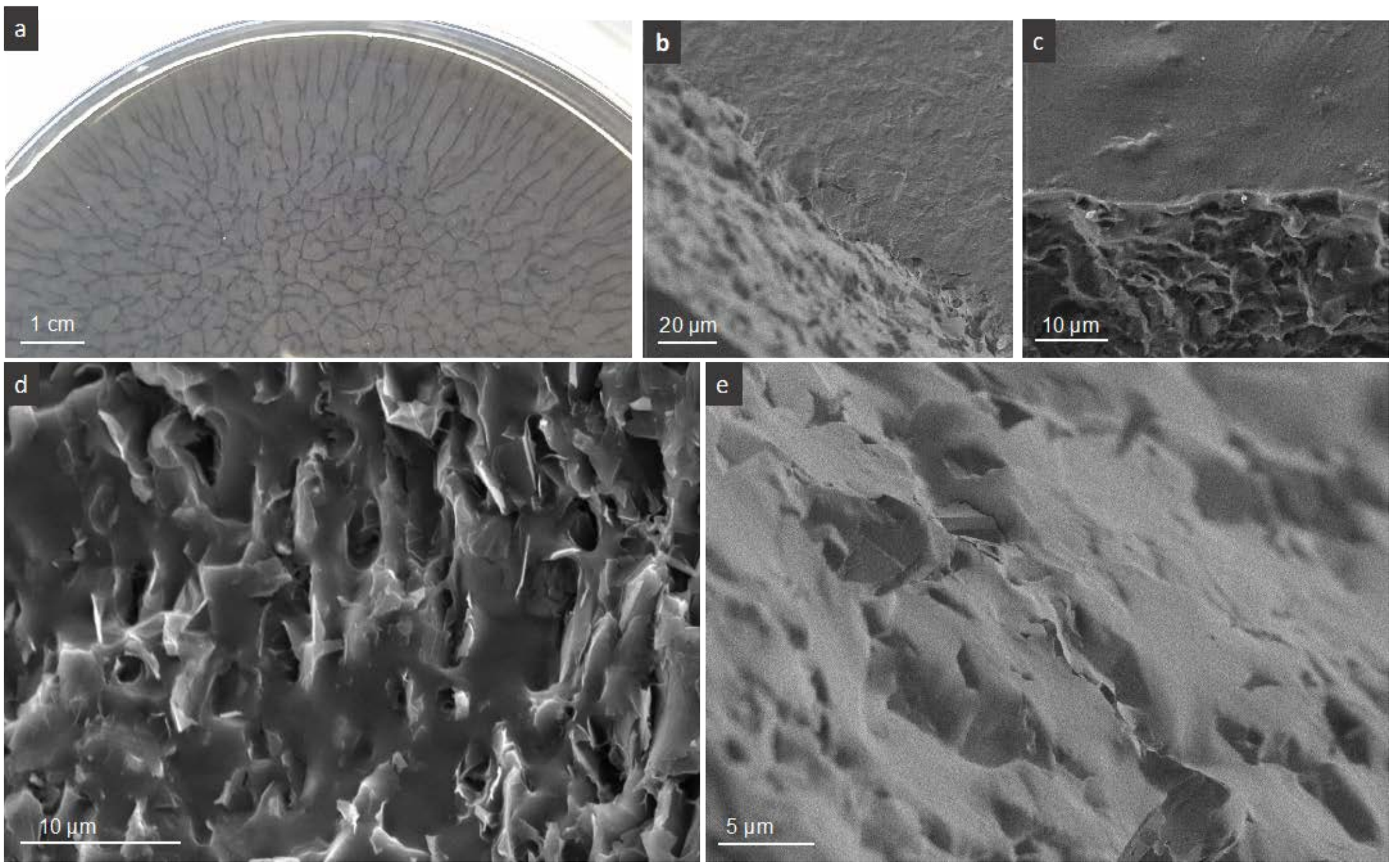

**Figure 2.** Representative PVA-FLG film structure, (**a**) optical photo of the drying film with formation of macroscopic branched network, (**b**,**c**) SEM micrographs of two sides of the films: opaque/upper and shining/down ones, (**d**,**e**) cross-sectional micrographs of the films: (**d**) closer to upper side, (**e**) closer to down side.

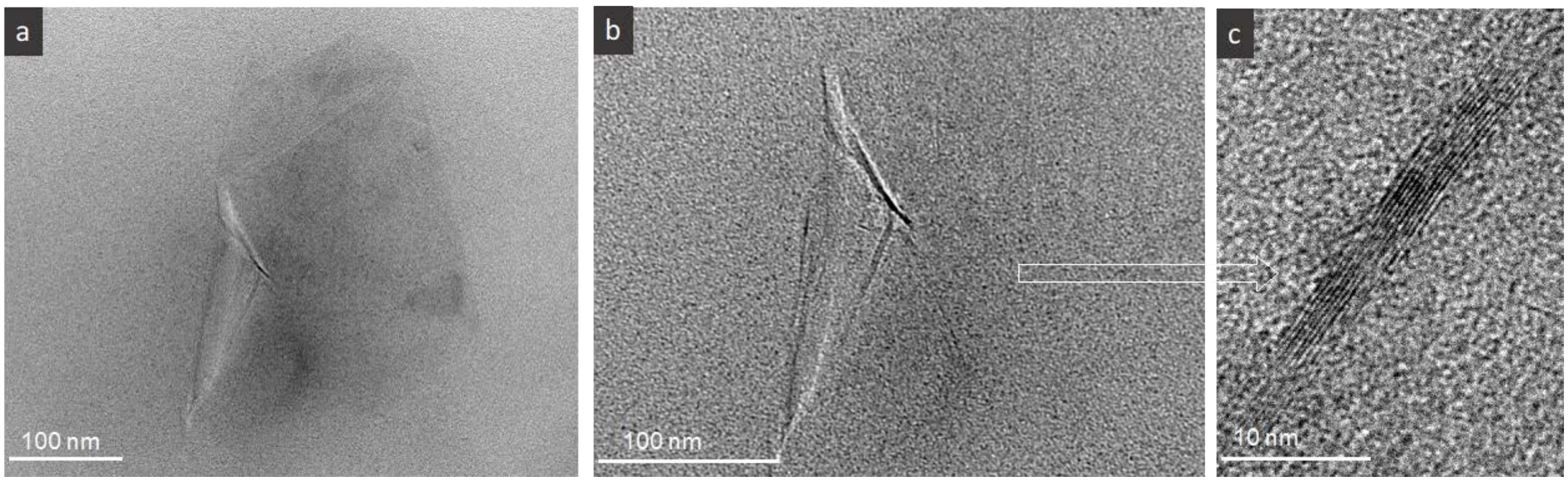

**Figure 3.** TEM micrographs of FLG flakes embedded in PVA matrix, (**a**) focus on the FLG morphology/size, (**b**) focus on the FLG edges (**c**) high resolution TEM of the edge of the flake demonstrating to have 9 sheets.

## *3.2. Thermal Properties of PVA-FLG*

The modification of the thermal stability of PVA films induced by the addition of FLG flakes is reflected in TGA analysis, series III, Figure 4. It seems that the weight loss at lower temperatures is more significant for the composites compared to PVA, while at higher temperatures, this tendency in the composites is less clear. The faster weight loss at the beginning in the composites can be first of all related to the presence of BSA, which as mentioned above has high hydrophilic lipophilic balance properties and can enhance the ability to lock up water molecules. In the last region of combustion, i.e., below 500 °C, the interface between FLG and remaining PVA slightly slows down the process. For more precise investigations, the combustion temperature at 10% and 50% weight loss ($T_{10\%}$, $T_{50\%}$) are shown in Table 1, Figure 4. As can be seen, along with the increase of FLG loading, $T_{10\%}$ decreases in a significant and consecutive manner. This tendency changes once all water and large part of BSA are gone, and at $T_{50\%}$, the most robust is the composite with 1% of FLG, then PVA and PVA-FLG 2% has similar thermal resistance, and only PVA-FLG 3% shows lower $T_{50\%}$. This random variation is probably linked to two phenomena, initially to the formation of a physical barrier from well dispersed FLG at 1% (decrease of the diffusion of degraded PVA to the gas phase) [29,30]. Next, at 3% FLG, to good thermal conductivity and related high heat transfer of FLG that accelerates a degradation of PVA.

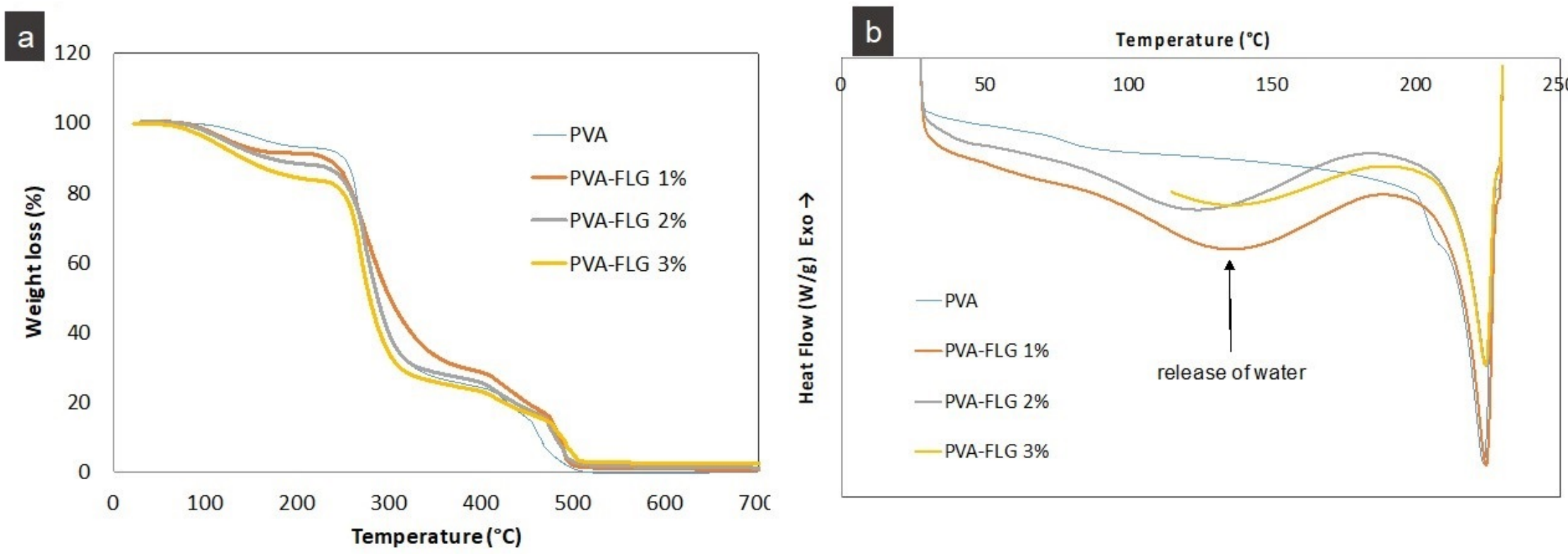

**Figure 4.** Thermal stability properties of the PVA-composites: (**a**) representative TGA (series III), (**b**) representative DSC (series IV).

**Table 1.** Temperatures for weight loss of 10 and 50% (series IV).

| Sample (Series IV) | $T_{10\%}$ | $T_{50\%}$ |
|---|---|---|
| PVA | 250 | 286 |
| PVA-FLG 1% | 228 | 300 |
| PVA-FLG 2% | 165 | 287 |
| PVA-FLG 3% | 141 | 278 |

The release of bound water, especially from the composites, is also clearly seen on DSC curves for the first heating, Figure 4b (series IV). According to DSC analysis, the melting temperatures ($T_m$) of the composites are in general slightly higher than $T_m$ of PVA (Table 2) but the calculated crystallinity degrees (%C) clearly decrease with addition of FLG. The crystallinities were obtained from melting enthalpies, $\Delta H_m$, values vs. $\Delta H_m$ of 100% crystalized PVA reference, i.e., 138.6 J/g [31], as shown in Table 2. The decrease of %C in the composites indicates disturbing effect of FLG additive on the ordering of the polymer chains and could be associated with so called hydrogen bond barrier blocking the hydrogen bonding within PVA. Such a hydrogen bond barrier was observed also in GO composites [7].

Likewise, the reduced melting enthalpy of the composites compared to the PVA in the first heat indicates loosely co-interacting species in the formers. This tendency is kept once water is released in the second heating only for certain samples. In few cases, the difference between relative maximum $\Delta H_{mII}$ obtained for the composites and reference PVA samples is much smaller, which can be attributed to the reduced mobility of the polymer chains interacting with BSA covered FLG. Some discrepancies in the modification of the relative $T_m$ and $\Delta H_m$ between the series are induced probably by coupled factors related to quantitative variation of the interactions between PVA, water, BSA-FLG and FLG edges considering variable size of FLG flakes. Despite these discrepancies, it can be observed that the crystallinity (and $\Delta H_{mII}$) in the second heating is relatively high for FLG at 1% and 2% loading compared to other composites and the reference samples.

Due to the locked-up water molecules the composites were additionally dried prior to the mechanical, conductive and barrier properties measurements.

**Table 2.** Melting temperatures and crystallinity degrees of PVA-FLG composites obtained from DSC analysis.

| | | $T_m$ [°C] | $\Delta H_m$ [J/g] | %C | $\Delta H_{m\ II}$ [J/g] |
|---|---|---|---|---|---|
| Series I | PVA | 224.1 | 66.8 | 48.8 | 93.3 |
| | PVA-FLG 0.5% | 224.6 | 63.3 | 45.7 | 75.6 |
| | PVA-FLG 1% | 224.4 | 66.2 | 47.7 | 78.7 |
| | PVA-FLG 3% | 224.68 | 58.9 | 42.5 | 73.2 |
| Series II | PVA | 220.6 | 49.9 | 36.0 | 87.5 |
| | PVA-FLG 0.3% | 224.6 | 54.5 | 39.3 | 79.6 |
| | PVA-FLG 0.5% | 220.6 | 47.1 | 34.0 | 63.1 |
| | PVA-FLG 1% | 224.1 | 61.9 | 44.7 | 71.4 |
| | PVA-FLG 2% | 224.8 | 67.6 | 48.8 | 87.4 |
| Series IV | PVA | 223.7 | 86.6 | 62.5 | 62.3 |
| | PVA-FLG 1% | 224.3 | 70.1 | 50.6 | 68.8 |
| | PVA-FLG 2% | 224.5 | 71.8 | 51.8 | 67.1 |
| | PVA-FLG 3% | 224.2 | 57.1 | 41.2 | 54.5 |

### *3.3. Mechanical Properties*

The investigated mechanical properties of the four composites series are presented in Figures 5 and 6a and Table 3. They include the representative stress-strain curves, the modification of tensile modulus as a function of FLG loading and enhancement of tensile modulus, strength and elongation at break. One can see in Figure 6a that the tensile modulus (E) varies in more or less significant manner between the series, while the optimum and critical loading of FLG in all series are the same, i.e., closed to 1%. The presented points are the average values while, taking into account the margins of error, we see that, for few series, optimum loading can be obtained at lower loading (around 0.5). Already, the addition of 0.3% wt. of FLG induces the enhancement of the properties, then the tensile modulus in each series increases with the increase of FLG amount until 1% of FLG and next it is strongly reduced demonstrating even poorer properties compared to the neat PVA. A similar tendency is maintained for the tensile strength. It seems that the improvement of these properties depends on the size of FLG flakes being the most significant in series III (the largest flakes), weaker in series II (the average size) with the smallest in series IV (the smallest flakes). The greatest improvement of E and tensile strength is measured for series III for 1% of FLG loading, 114% and 60% respectively. Series I, containing also large flakes, shows the same tendency but the enhancement is somehow weaker. (Let recall that the drying step of films in series I was much faster, occurring at higher temperature, then drying of other series.) The elongation at break diminishes with the addition of FLG and this behavior is also measured in series III. On the contrary, the elongation at break increases with addition of FLG in series II and especially in series IV, so, inversely to FLG size. The superior elongation at break properties of series II and IV containing smaller sheets, FLG-4h and FLG-6h, indicate that the interactions between FLG and PVA occur mostly via the FLG edges and the impact of the adsorbed over FLG surface albumin on the overall interactions is lower. Consequently, the gliding of the FLG sheets is easier compared to the other series. On the contrary, lower ductility along with superior tensile modulus and strength in series with bigger FLG sheets suggests a relatively higher interaction through the bonding between PVA and BSA adsorbed over FLG surface. The potential impact of the FLG size on the formed branched patterns and consequently on the mechanical properties cannot indeed be excluded and would require separate investigations, while initial observations suggest no remarkable modification of the pattern between the series.

Our results are to a certain extent in agreement with the ones reported earlier where the best results were obtained for the PVA composites containing near 1% of GO and rGO (0.8%). Likewise, the enhancement of GO or rGO content to 1.6 resulted in the reduction of mechanical properties [7]. The composites prepared by Zhao behaved in a similar way, the mechanical percolation was obtained for 1.8% [8]. We can say then, that in this work, despite the formation of branched FLG pattern, the optimum percolation is obtained at similar FLG content range to the one obtained with GO/rGO PVA films.

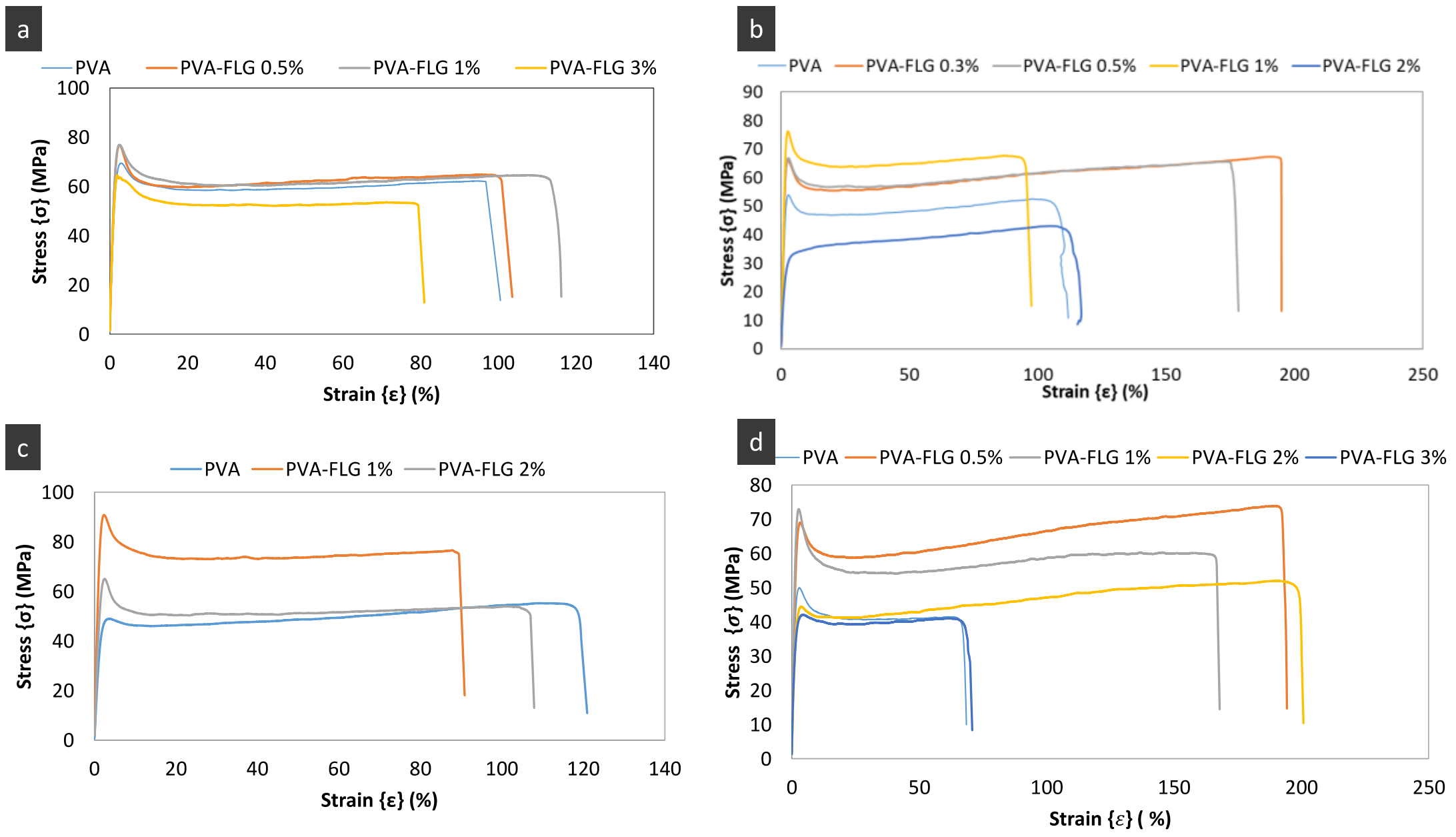

**Figure 5.** The representatives stress-strain curves of PVA-FLG composites: (**a**) series I, (**b**) series II, (**c**) series III, (**d**) series IV.

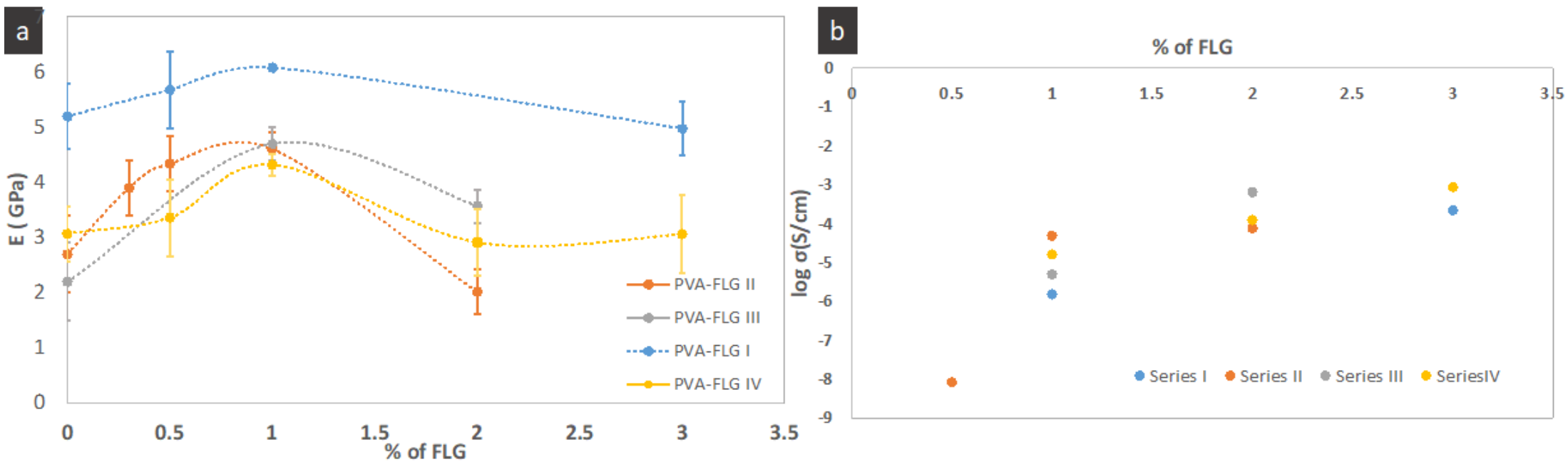

**Figure 6.** (**a**) Young modulus of PVA-FLG composites as function of FLG %, (**b**) conductivity of PVA-FLG composites as a function of FLG%.

**Table 3.** Enhancements of mechanical properties in PVA-FLG composites.

| | PVA-FLG (%) | Enhancement of Tensile Modulus (GPa) (%) | Enhancement of Tensile Strength (MPa) (%) | Enhancement of Elongation at Break (%) (%) |
|---|---|---|---|---|
| Series I | 0.5 | 9 | 2 | −6 |
| | 1 | **22** | **2** | 17 |
| | 3 | −4 | −16 | −20 |
| Series II | 0.3 | 44 | 18 | 52 |
| | 0.5 | 61 | 17 | 9 |
| | 1 | **70** | **29** | 22 |
| | 2 | −25 | −17 | 9 |
| Series III | 1 | **114** | **60** | −36 |
| | 2 | 62 | 41 | −16 |
| Series IV | 0.5 | 9 | **43** | **73** |
| | 1 | **42** | **39** | **75** |
| | 2 | −5 | 24 | **67** |
| | 3 | −0.3 | 0 | −30 |

Since PVA is a semicrystaline polymer, its mechanical properties should also be affected by crystallinity features. Indeed, in series IV, the melting temperature slightly increases in the nanocomposites compared to PVA, while the crystallinity (from the second DSC heat rate) in the composites with 1% and 2% FLG loading is even higher than for the reference PVA.

Taking all these into account, the improvement of the mechanical properties in the composites at FLG loading around 1% wt. can then be attributed to the enhanced crystallinity, interactions between FLG-BSA/FLG-edges and PVA chains through hydrogen bonding, and van der Waals interactions as described above.

### 3.4. Electrical Conductivity

According to the FPPs measurements, the films are conductive, starting from 0.5% of FLG content with conductivity order of $10^{-9}$ S/cm, as shown in Figure 6b. The conductivity strongly increases, reaching values in the order of $10^{-6}$–$10^{-5}$ S/cm at 1% FLG loading and next increases slowly to be the order of $10^{-5}$–$10^{-4}$ S/cm at 2% FLG loading, being finally stabilized with the order of $10^{-4}$ S/m for FLG content of 3%. The highest value, almost $10^{-3}$ S/cm of order ($8.5 \times 10^{-4}$ S/cm), was achieved for series IV with 3% of FLG content.

A similar conductivity enhancement behavior was observed for rGO based composite, where the conductivity clearly jumped going from 0.8% to 1.6% of rGO and then stabilized at higher rGO concentration [7].

We assume that the inhibited increase of conductivity at certain FLG concentration is related to the increasing face-to-face arrangement of the FLG flakes. Such overlapping of the flakes in z-direction increases more-and-more along with FLG content. The conductivity in the direction perpendicular to the plane is three orders of magnitude lower than the planar one in FLG flake as it is in graphite [20], and an additional resistance in the films occurs at the connections of the flakes. Likewise, some extra overlapping of the flakes at higher FLG concentration can be induced by the formation of branched network, instead of homogenous distribution of add-FLG in the plane. Taking into account the limit of pattern formation (or random distribution) of FLG in the x direction, the obtained conductivities are relatively high. The conductivities and percolation thresholds of the composites are within the highest and the lowest ones, respectively, known in the literature for PVA-graphene based films [10,14]. Comparing our results with the ones achieved for PVA-rGO film, the PVA-FLG composites show around 5 orders of magnitudes higher conductivities for the same content of carbon [7]. The measured here conductivities are also comparable with the ones obtained for the composites containing large rGO of few tens of macrons in majority [10]. The effect of the aspect ratio on the conductivity properties and more precisely on the percolation threshold has been clearly shown for rGO flakes [10]. Indeed, we highlight that four main factors impact the transport properties in the films: carbon crystallinity, carbon morphology, carbon arrangement/interactions, and carbon-PVA interactions.

We did not find any specific correlation between the size of FLG sheets and conductivity percolation or behavior in general, but the differences in FLG size and % loading are possibly too small to observe a conductivity–size relationship. Likewise, the formation of the branched network would modify this simple reliance.

The good conductive characteristic despite a relatively significant thickness of the flakes is first of all ensured by the well crystalized carbon lattice of FLG with a low amount of defects and relatively large lateral size of the flakes compared to other FLG materials obtained by the liquid exfoliation of graphite [21]. In terms of FLG arrangement, there are two main factors here that play a role, i.e., the formation of connected FLG branched macroscopic patterns and the flat arrangement of the flakes at the conductive side as discussed next. Having seven graphene sheets on average and comparable or lower to rGO lateral size of the sheets, the FLG used here would make percolation a few times higher in terms of concentration than the thin graphene (rGO) if the advantages of the arrangement are not present. It was shown previously that the formation of fractal-like networks formed by pure FLG flakes over glass substrate enhanced the transport properties over macroscopic substrate compared to random distribution, but this effect in herein composites is for now difficult to determine [28]. Regardless the macroscopic network, we need to note the fact that, in all measured films, only the opaque side of the films is conductive—the one that initially was in the contact with glass substrate during the casting of the films, as shown in Figure 2c. This was quite surprising taking into account that, according to the first SEM observations, the presence of FLG flakes is clearly pronounced in the upper side of films. Despite relatively rapid drying and quite viscous colloids, we could suppose that some low difference in % loading exists between the two sides. For this purpose, the XPS analysis on the surface of both sides of the film was performed and did not reveal any chemical difference (the relative C to O ratio would show this difference if existing). This means that the concentration in FLG of both sides is very close while morphology can be side-dependent. This latter was next confirmed by applying high electron intensity mode under SEM analysis.

The low and higher magnifications micrographs on Figure 7 demonstrate two sides of the film where, due to the charging effect the FLG, flakes and PVA domains can be clearly distinguished as black and white, respectively. It is clear that, at the conductive side, we see a quasi-totally flat arrangement of the flakes that results in their good connectivity making impression of FLG excess, Figure 7a,b. Such an arrangement is disturbed at the opposite, upper side, through which the evaporated water molecules and air are evacuated upon drying living space for FLG flakes collapsing. On this side, we see more protruded flakes with their edges then the flakes surface as it is the case of the conductive side, this time giving impression of PVA excess. This side view is in agreement with the above cross-sectional observations, Figure 2d. It results in a lack of FLG connection and conductivity.

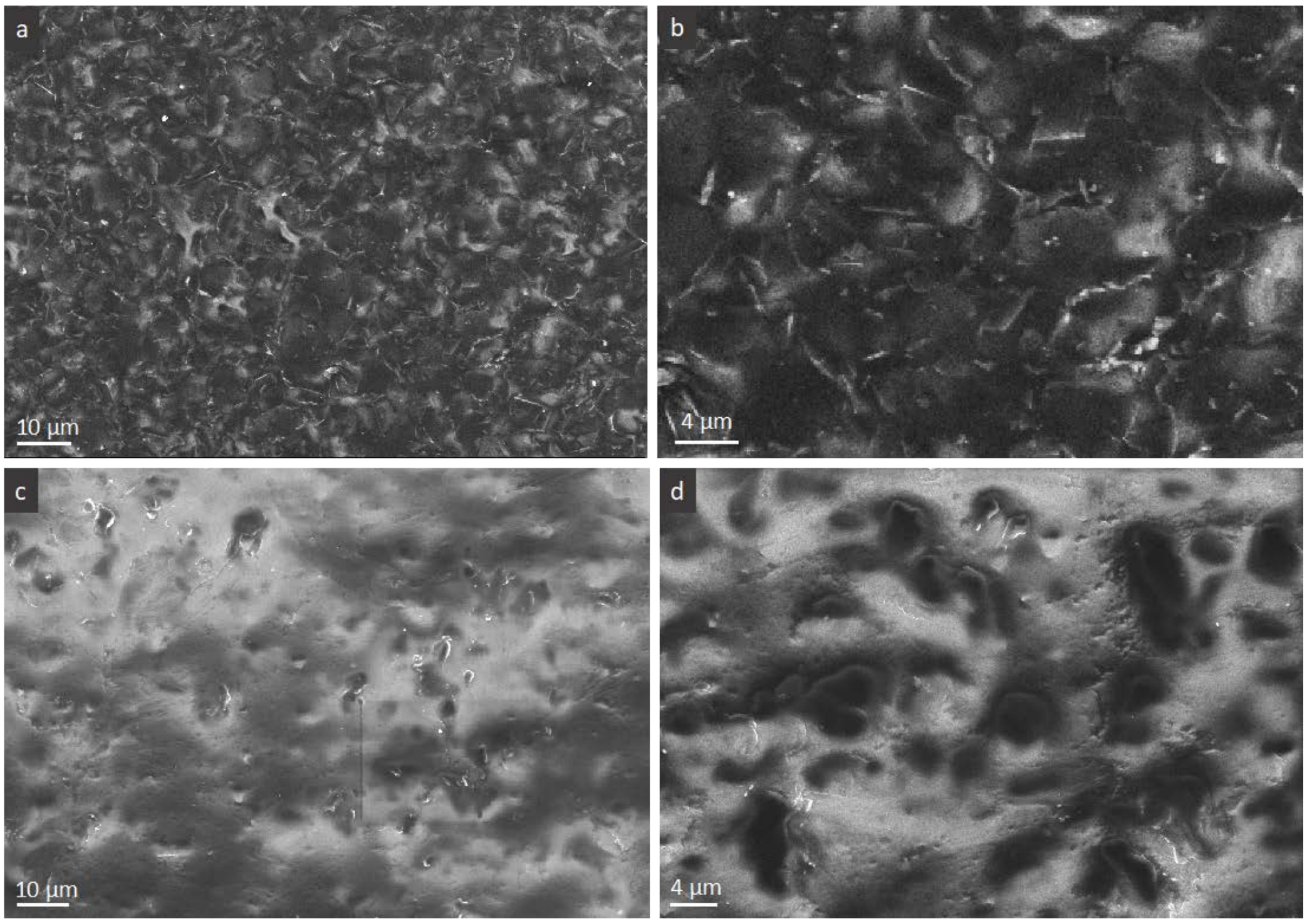

**Figure 7.** SEM micrographs of PVA-FLG films, series IV, 3% of FLG (high electron intensity): (**a**,**b**) conductive side, (**c**,**d**) non-conductive side.

### 3.5. Oxygen Barrier Properties

Few of prepared composites from series I and II with different FLG % were submitted to the oxygen barrier properties measurements. The oxygen transmission decreases up to 60%, from 10.7 for pure PVA to 4.3 cc/m$^2$ day for series I-0.5%FLG, and close to this, to 4.5 and 4.7 cc/m$^2$ for series II-0.3%FLG and series I-0.5%FLG. These average barrier properties are in agreement with the non-homogenous distribution of and specific arrangement of FLG into connected macroscopic network. It seems also that, for higher concentration of FLG, the composites show slightly higher transmission (4.9–6.1 cc/$m^2$), which would also suggest that the successive addition of FLG allows added overlapping of FLG in the z direction within the network than the extension of the network in the plane.

## 4. Conclusions

PVA composites containing FLG additive show enhanced mechanical, conductive, and gas barrier properties. The best mechanical properties are observed for PVA containing 1% wt. of FLG. The most significant enhancements of tensile modulus and strength are measured for the composites containing larger FLG flakes contrary to the elongation at break for which the best improvement is noticed for PVA containing lower size flakes. The rapid, simple, bio-compatible, and efficient synthesis of FLG together with the resulting large size of FLG sheets make this FLG material a suitable choice for further applications and their optimization in polymer composites. Preliminary studies show that highly conductive polymer-FLG films can be obtained, while at present, the formation of a macroscopic branched FLG network and one-side conductivity are achieved.

**Supplementary Materials:** The following are available online at http://www.mdpi.com/2079-4991/10/5/858/s1, Figure S1: SEM micrograph of FLG flakes with adsorbed BSA, Figure S2: Optical image of PVA film, Figure S3: SEM micrograph of the surface of PVA-FLG film (a branch of network).

**Author Contributions:** Synthesis of FLG and composites, characterization of composites (TGA, mechanical tests, SEM), B.V.d.S.; DSC, conductivity measurements, H.E.M.; characterization of FLG (XPS, TGA, SEM), A.M.; conductivity measurements, P.L.; DSC characterization, C.S.; SEM microscopy, T.R.; principal investigator/project coordinator, supervision, writing, I.J. All authors have read and agreed to the published version of the manuscript.

**Funding:** This work was supported by Centre National de la Recherche Scientifique (CNRS), pre-maturation program.

**Acknowledgments:** The PCA accredited laboratory No. AB 1450, located at the Centre from Bioimmobilisation and Innovative Packaging Materials in Szczecin (Poland) is acknowledged for $O_2$ transmission measurements. Thierry Dintzer (ICPEES) are acknowledged for high electron intensity SEM microscopy measurements. Dris Ihiawakrim (IPCMS) is acknowledged for TEM microscopy, Damien Favier (ICS) is acknowledged for help with traction tests.

**Conflicts of Interest:** The authors declare no conflict of interest.